\documentclass[12pt]{article}

\usepackage{graphicx}
\usepackage{epstopdf}
\usepackage{amssymb}

\def\be{\begin{equation}}
\def\ee{\end{equation}}
\def\bea{\begin{eqnarray}}
\def\eea{\end{eqnarray}}

\def\tev{\, {\rm TeV}}
\def\gev{\, {\rm GeV}}

\newcommand{\gsim}{\lower.7ex\hbox{$\;\stackrel{\textstyle>}{\sim}\;$}}
\newcommand{\lsim}{\lower.7ex\hbox{$\;\stackrel{\textstyle<}{\sim}\;$}}

\newcommand{\pb}{\rm pb}

\newcommand{\Dsl}[1]{\slash\hskip -0.20 cm #1}

\newcommand\pubnumber{UH-511-1230-14}
\newcommand\pubdate{\today}

\def\napoli{Department of Physics and Astronomy, \\
University of Hawaii, Honolulu, HI 96822 USA}

\def\Title#1{\begin{center} {\Large #1 } \end{center}}
\def\Author#1{\begin{center}{ \sc #1} \end{center}}
\def\Address#1{\begin{center}{ \it #1} \end{center}}

\newcommand\pubblock{\rightline{\begin{tabular}{l} \pubnumber\\
         \pubdate  \end{tabular}}}
\newenvironment{Abstract}{\begin{quotation}  }{\end{quotation}}
\newenvironment{Presented}{\begin{quotation} \begin{center} 
             PRESENTED AT\end{center}\bigskip 
      \begin{center}\begin{large}}{\end{large}\end{center} \end{quotation}}
\def\Acknowledgements{\bigskip  \bigskip \begin{center} \begin{large}
             \bf ACKNOWLEDGMENTS \end{large}\end{center}}

\def\beq{\begin{equation}}
\def\eeq#1{\label{#1}\end{equation}}
\def\eeqn{\end{equation}}

\def\beqa{\begin{eqnarray}}
\def\eeqa#1{\label{#1}\end{eqnarray}}
\def\eeqan{\end{eqnarray}}

\let\bar=\overbar

\def\Dslash{\not{\hbox{\kern-4pt $D$}}}
\def\dslash{\not{\hbox{\kern-2pt $\del$}}}

\def\ee{e^+e^-}

\def\msb{{\bar{\ssstyle M \kern -1pt S}}}

\begin{document}
\begin{titlepage}
\pubblock

\vfill
\Title{WIMPy Leptogenesis}
\vfill
\Author{ Patrick Stengel}
\Address{\napoli}
\vfill
\begin{Abstract}
We consider a class of leptogenesis models in which the lepton asymmetry
arises from dark matter annihilation processes which violate $CP$ and
lepton number.  Importantly, a necessary one-loop contribution
to the annihilation matrix element arises from absorptive final state interactions.
We elucidate the relationship between this one-loop contribution and
the $CP$-violating phase.
As we show, the branching fraction for dark matter
annihilation to leptons may be small in these models, while still generating the necessary
asymmetry.
\end{Abstract}
\vfill
\begin{Presented}
CosPA 2013 \\
Honolulu, HI,  November 12--15, 2013
\end{Presented}
\vfill
\end{titlepage}
\def\thefootnote{\fnsymbol{footnote}}
\setcounter{footnote}{0}

\section{Introduction}

Both the baryon asymmetry of the universe (BAU) and the existence of non-baryonic dark matter (DM)
are well motivated by observation at a variety of scales and epochs. Recent observations of the cosmic microwave background (CMB) by the Wilkinson Microwave Anisotropy Probe (WMAP)~\cite{WMAP:2011} establish the baryonic and cold
 dark matter densities in our universe as
\bea
\Omega_b h^2 &\sim & 0.022  ,
\nonumber\\
\Omega_{DM} h^2 &\sim & 0.12 .
\eea
The most common theoretical paradigm used to explain the factor of $ \sim 5 $ between the two observed densities has been asymmetric dark matter (ADM) models~\cite{Zurek:2013}. While ADM models transfer an asymmetry of the dark sector into the
baryonic sector, dynamical generation of the baryon asymmetry from dark matter decays have also been investigated~\cite{Kohri:2009}. More recently, weakly interacting massive particle (WIMP) annihilation has been suggested as
a mechanism for baryogenesis~\cite{Cui:2012}. In such WIMPy models, a departure from thermal equilibrium is guaranteed by
the WIMP framework. As such, these models of DM annihilation need only violations of baryon number, $C$ and $CP$ in order to
satisfy the Sakharov conditions for baryogenesis~\cite{Sakharov:1967dj}.     

\subsection{Matter Asymmetry}

In order to demonstrate the mechanism of $CP$ violation when dynamically generating an asymmetry, consider a generic  DM
annihilation to a multi-particle final state, $XX \rightarrow Y$, and the $CP$ conjugate process $XX \rightarrow \bar Y$.
We can write the matrix element for such processes as the sum of $CP$-invariant and $CP$-violating terms,

\bea
{\cal M}_{XX \rightarrow Y} &=& {\cal M}_{XX \rightarrow Y}^{CP} + {\cal M}_{XX \rightarrow Y}^{CPV} ,
\nonumber\\
{\cal M}_{XX \rightarrow \bar Y} &=&\pm \left( {\cal M}_{XX \rightarrow Y}^{CP} - {\cal M}_{XX \rightarrow Y}^{CPV} \right) .
\eea

As a result of the optical theorem, the relative phase between the $CP$-invariant and $CP$-violating terms will be $\pm \pi /2$ at tree
level, eliminating any asymmetry. Thus, any nonvanishing asymmetry in
the production cross section must be an interference of $CP$-invariant and $CP$-violating matrix
elements with a relative phase difference from an additional imaginary matrix element contribution at 
loop level,

\bea
\sigma_{XX \rightarrow Y} - \sigma_{XX \rightarrow \bar Y} &\propto & Re \left[{\cal M}_{XX \rightarrow Y}^{CP}
({\cal M}_{XX \rightarrow \bar Y}^{CPV} )^* \right].
\eea

In standard models of baryogenesis the relative phase is introduced by the interference of a $CP$-invariant tree-level
diagram with a $CP$-violating loop-level diagram. Alternatively, we separate the $CP$-violating phase of two tree-level diagrams
from an absorptive phase introduced by a final state decay. 

\subsection{WIMPy Baryogenesis Models}

Before discussing a particular WIMPy model, some general features and issues are worth noting.
A pair of WIMPs, stabilized by a discrete symmetry, will annihilate to a quark (or lepton) and a heavy exotic field
through a set of operators with the $C$ and $CP$ properties necessary to satisfy the Sakharov conditions.
This new state needs to also be protected from interactions with standard model by a discrete charge and should transfer its own asymmetry into a hidden sector. In models with a relative phase generated by the interference of tree-level and
loop-level diagrams, dangerous washout diagrams arise. Washout processes yield
terms in the Boltzmann equations for the rate of the asymmetry production that are proportional to the asymmetry itself. As a consequence, these $CP$-invariant terms will dampen the generation of an asymmetry.

\begin{figure}[htb]
\centering
\includegraphics[width=\textwidth]{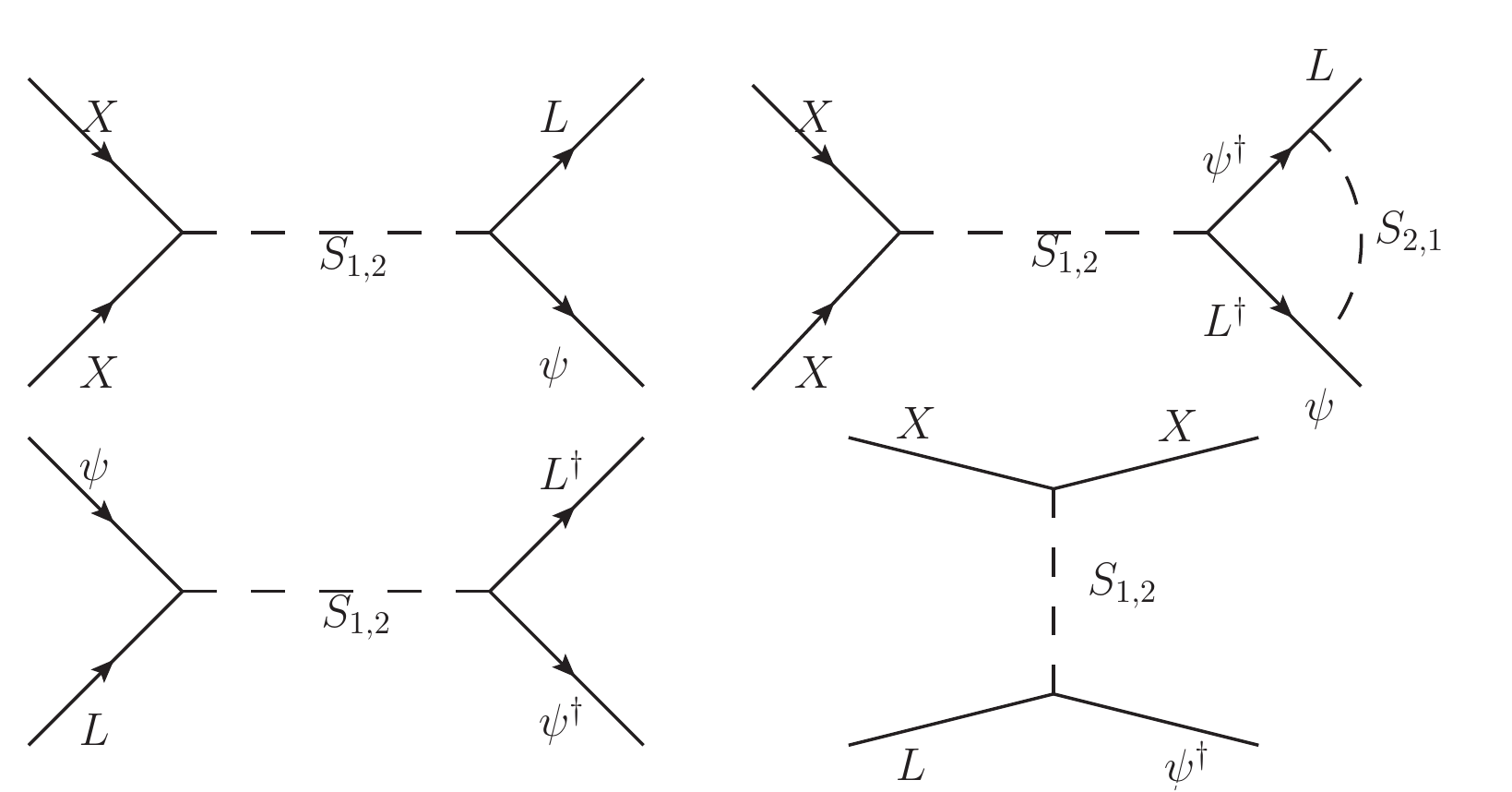}
\caption{Asymmetry generating DM, $X$, annihilation diagrams and corresponding washout diagrams in WIMPy models~\cite{Cui:2012}. Note that in order to generate a non-vanishing asymmetry, DM must annihilate to a SM lepton (or quark), $L$, and an additional weak scale field, $\psi$, through the interference of tree-level (top left) and loop-level (top right) diagrams. The loop diagram implies the existence of the leading order, ``pure", washout process (bottom left) and crossing the tree-level DM annihilation implies the ``mixed" washout diagram (bottom right). }
\label{fig:loops}
\end{figure}

The presence of these dangerous washout processes are clearly implied by the loop-level DM annihilation diagram in Figure~\ref{fig:loops}. Because there is
no dark matter in the initial state for these ``pure" washout processes, they lack the Boltzmann suppression present in ``mixed" washout processes. Thus, the $CP$-violating processes must account for a large portion of the total DM annihilation in order to generate a large enough asymmetry. As the mediator, $S_{1,2}$, runs in the loop, the cross section of the asymmetry generating process will also be inversely proportional to the square of the mediator mass. The result is a necessarily small separation of scales between the new external states, $X$ and $\psi$, and the mediator $S_{1,2}$. This is generally problematic for models constructed within the framework of an effective field theory (EFT) and for WIMPy models which attempt to independently resolve the DM relic density, BAU and ultraviolet (UV) physics.

\section{Field Content and Leptogenesis Model}

\begin{table}[hear]
\centering
\begin{tabular}{|c|c|c|c|c|}
\hline
 Fields & $SU(2)_L$ &  $ Q_{U(1)_Y} $  & $ Q_{U(1)_L}$ & $\mathbb {Z}_2$ \\
\hline
$X$ & 1 & 0 & 0 & - \\
\hline
$P_L L = l_L$ & $\Box$ & -1/2 & +1 & +  \\
$H$ & $ 1 $ & 0 & 0 & + \\
\hline
$P_L L = \nu_L$ & $\Box$ & -1/2 & +1 & +  \\
$H$ & $ 1 $ & 0 & 0 & + \\
\hline
\end{tabular}
\caption{Particle Content}
\label{tab:particles}
\end{table}

We propose a generalized EFT containing generic WIMPy field content
\cite{Cui:2012, Bernal:2012}. The DM, $X$, and new heavy exotic, $H$, are protected by charges shown in
Table~\ref{tab:particles}. Unlike previous work, we don't assume $H$ is prevented from possibly large washout terms (through gauge interactions) by the same discrete symmetry as $X$. For simplicity, we assume the DM is protected by a $Z_2$ and annihilates to $H$ and a SM lepton, $L$, through operators which violate $SU(2)_L$
and, thus, the EFT couplings are proportional to a Higgs vev.

\bea
{\cal O}_1 &=& {\lambda_1 \over 2M_*^2} (\imath \bar X \gamma^5 X) (\bar H P_L L)
+ {\lambda_1^* \over 2M_*^2} (\imath \bar X \gamma^5 X) (\bar L P_R H)
\nonumber\\
{\cal O}_2 &=& {\lambda_2 \over 2M_*^2} (\bar X \gamma_\mu \gamma^5 X) (\bar H \gamma^\mu P_L L)
+ {\lambda_2^* \over 2M_*^2} (\bar X \gamma_\mu \gamma^5 X) (\bar L \gamma^\mu P_L H )
\label{eq:operator}
\eea

A model in which both $X$ and $H$ are protected by a $Z_4$ is phenomenologically very similar, only the DM must be Dirac due to the imaginary charges then assigned to $X$. Note the DM bilinears in these dimension six operators are the only combination which can yield $CP$-violating interference terms for
Majorana fermion DM~\cite{Kumar:2013iva}. Since ${\cal O}_1$ is $CP$ odd if $\lambda_1$ is real and ${\cal O}_2$ is $CP$ odd if $\lambda_2$ is imaginary, a non-vanishing $CP$-violating phase will arise from a term proportional to 
$Re( \lambda_1 \lambda_2^* )$.

 \subsection{Absorptive Final State Decays}

In order to generate an asymmetry given our $CP$-violating phase, we also need an absorptive phase from a final state $H$ decaying into light hidden sector fields, scalar $\phi$ and fermion $H'$. The loop, typically manifested by final state vertex corrections in standard leptogenesis scenarios, is alternatively provided by the external leg correction of the unstable $H$. The fully corrected propagator for a fermioninc intermediate state is given by~\cite{Kniehl:2008cj}:

\bea
S(\Dsl p) = {{\Dsl p}_H + (m_H - \imath \Gamma_H /2) \over p_H^2 - m_H^2 -\imath m_H \Gamma_H }.
\eea

If the width of the $H$ resonance is small there will be regions of phase space where the intermediate $H$ goes on-shell, generating an asymmetry between the cross sections for $XX \rightarrow \phi^* \bar H' L$ and $XX \rightarrow \phi \bar L  H'$. As a
consequence of the optical theorem, we can see the imaginary matrix element contribution we need to preserve our $CP$ asymmetry can only come from a term
proportional to $ \Gamma_H / 2 $ in the propagator. If we assume both $ \lambda_1$ and $ \lambda_2$ are real, ${\cal O}_1$ will
maximally violate $CP$ and only couple to the right-handed Weyl spinor $H_R$, while ${\cal O}_2$ will
completely preserve $CP$ and only couple to the left-handed Weyl spinor $H_L$. We assume
$H$ can only decay from the left-handed helicity (as in the standard model) into our hidden sector through the $CP$-invariant operator

\bea
{\cal O}_H &=& |g| (\phi^* \bar H' P_L H + \phi \bar H P_R H').
\eea

Thus the relative phase between the $CP$-violating $XX \rightarrow \bar H_R L_L \rightarrow \phi \bar H'_R  L_L$ and $CP$-conserving $XX \rightarrow \bar H_L L_L \rightarrow \phi \bar H'_R  L_L$ must arise from the $-\imath \Gamma_H /2$ contribution to the helicity-flip term of the propagator. Assuming the narrow width approximation, we can then calculate the asymmetry in the full cross sections of the $2 \rightarrow 3$ DM annihilation processes,

\bea
(\sigma^{XX \rightarrow \phi^* \bar H' L } - \sigma^{XX \rightarrow \phi \bar L H' } ) v &=&
\Gamma_H   {  Re( \lambda_1 \lambda_2^* ) m_X \over  4\pi M_*^4 }
\left[1- {m_H^2 \over s }\right]^2 .
\eea

Note $\sqrt{s}$ is the energy in center-of-mass frame and for simplicity we assume $m_{H'}, m_\phi, m_L \ll m_X , m_H \sim \tev $. In order to keep our UV mediator effectively decoupled from low-energy physics, we set $M_* = 10~\tev$. If we define a measure of the asymmetry,

\bea
\epsilon &\equiv& {\sigma^{XX \rightarrow \phi^* \bar H' L } - \sigma^{XX \rightarrow \phi \bar L H' }
\over \sigma^{XX \rightarrow \phi^* \bar H' L } + \sigma^{XX \rightarrow \phi \bar L H' } },
\eea

then we find $\epsilon \sim \Gamma_H / m_{H,X}$. This is in contrast with other WIMPy baryogenesis models, where
one typically finds $\epsilon \sim m_X^2 / M_*^2$~\cite{Bernal:2012} due to the $CP$-violating loop diagrams. The consequences of this distinction in our model are explained in the next section.

\subsection{Boltzmann Equations}

We can write the Boltzmann equations in terms of dimensionless variables $ x = m_X / T$ and $Y = n / s $,
where $n$ is the number density and $s$ is the entropy density. Assuming an adiabatic process, the entropy $S$ should be constant,
and $Y$ is essentially a comoving number density. As the hidden sector fields, $\phi$ and $H'$, are effectively massless, they should remain approximately in equilibrium throughout the relevant cosmological epoch. $L$ is also light but we want to track even a small departure from equilibrium in order to generate the small observed baryon (or, through sphalerons, lepton) asymmetry
$Y_{\Delta L} \equiv Y_L - Y_{\bar L} \sim 10^{-10} $. Thus, we can assume $Y_L + Y_{\bar L} \simeq 2Y_{L_{eq}}$. The coupled Boltzmann equations for the primary DM annihilation to any fermion/antifermion pair, as well as the subdominant DM annihilation that gives us the injection of our lepton asymmetry are~\cite{Kolb:1990}:

\bea
{x^2 H(m_X) \over s(m_X) } {dY_X \over dx}  &=& - \langle \sigma_A v \rangle (Y_X^2 - Y_{X_{eq}}^2 ) ,
\label{eq:BoltzmannDM}
\\
{x^2 H(m_X) \over s(m_X) } {dY_{\Delta L}^{inj} \over dx}  &=&  {1 \over 2}[ \langle \sigma_{XX \rightarrow
\phi^* \bar H' L } v \rangle  ]
(Y_X^2 - Y_{X_{eq}}^2 Y_L / Y_{L_{eq}} )
\nonumber\\
&\,& -{1 \over 2}[ \langle \sigma_{XX \rightarrow \phi \bar L H' } v \rangle  ] (Y_X^2 - Y_{X_{eq}}^2 Y_{\bar L }/ Y_{\bar L_{eq}} )
\nonumber\\
&\,&
-\langle \sigma_{XL \rightarrow \phi X H' } v \rangle Y_X (Y_{L} - Y_{L_{eq}} )
+\langle \sigma_{X\bar L \rightarrow \phi^* X \bar H' } v \rangle Y_X (Y_{ \bar L} -Y_{\bar L_{eq}})
\nonumber\\
&\,& +...
\label{eq:BoltzmannAsym}
\eea

The ``$+...$" terms involved suppressed processes in which an on-shell resonance is kinematically forbidden.
In order to achieve the correct relic DM density, given a weak scale $m_X$, we always set $ \langle \sigma_A v \rangle = 1~\pb$.
Note that $H(T)$ is the Hubble parameter at temperature $T$ given a flat, radiation-dominated early universe. While the 
equilibrium rates for the $ 3 \rightarrow 2 $ processes are given by detailed balance with the rates for the $ 2 \rightarrow 3 $ processes, the actual $ 3 \rightarrow 2 $ rates are calculated by rescaling the equilibrium rates by the ratio of the actual mass densities to the equilibrium mass densities. We can rewrite the the equation for $dY_{\Delta L}^{inj} / dx$ as

\bea
{x^2 H(m_X) \over s(m_X) } {dY_{\Delta L}^{inj} \over dx}  &\sim& \langle \sigma_{XX}^{CPV} v \rangle (Y_X^2 - Y_{X_{eq}}^2)
- \langle \sigma_{XX}^{CP} v \rangle Y_{X_{eq}}^2 Y_{\Delta L} / Y_{L_{eq}}
- \langle \sigma_{XL}^{CP} v \rangle Y_X Y_{\Delta L} ,
\nonumber
\eea
where
\bea
\langle \sigma_{XX}^{CP} v \rangle &\equiv & {1\over 2} \left[
\langle \sigma_{XX \rightarrow \phi^* \bar H' L } v \rangle + \langle \sigma_{XX \rightarrow \phi \bar L H' } v \rangle \right] ,
\nonumber\\
\langle \sigma_{XX}^{CPV} v \rangle &\equiv & {1\over 2} \left[
\langle \sigma_{XX \rightarrow \phi^* \bar H' L } v \rangle - \langle \sigma_{XX \rightarrow \phi \bar L H' } v \rangle \right] ,
\nonumber\\
\langle \sigma_{XL}^{CP} v \rangle &\equiv & {1 \over 2} \left[
\langle \sigma_{XL \rightarrow \phi X H' } v \rangle + \langle \sigma_{X\bar L \rightarrow \phi^* X \bar H' } v \rangle \right] .
\label{eq:SimplifiedCrossSection}
\eea

Before the electroweak phase transition (EWPT) any asymmetry generated within leptons will be shared with baryons through electroweak sphalerons. Recent discovery of the Higgs mass allows for lattice calculation of the sphaleron rate through the EWPT until sphalerons are effectively decoupled~\cite{D'Onofrio:2012}.
The expression for the lepton injection rate is the source term in the coupled differential equations for the evolution of the baryon and lepton number densities.

\bea
x H(T) {dY_{\Delta B} \over dx} &=& - \gamma (T) \left[ Y_{\Delta B} + 3 \eta (T) Y_{\Delta L} \right]
\nonumber\\
x H(T) {dY_{\Delta L} \over dx} &=&  - {1 \over 3} \gamma (T) \left[ Y_{\Delta B} + \eta (T) Y_{\Delta L} \right]
+ x H(T) { dY_{\Delta L}^{inj} \over dx}
\eea   

\begin{figure}[hear]
\center\includegraphics[width=\textwidth]{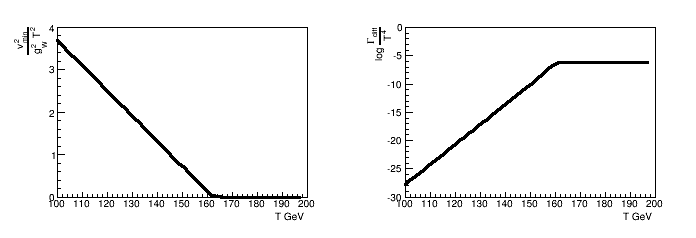}
\caption{Lattice calculation of $v_{min}^2 / (T^2 g_{weak}^2)$ (left) and $\log (\Gamma_{diff} /T^4)$
(right) through the transition region~\cite{D'Onofrio:2012}.  For $T<140~\gev$, $\log (\Gamma_{diff} /T^4)$ is
extrapolated from analytical calculations deep in the broken phase~\cite{Burnier:2006}, with a constant
rescaling to provide consistency with the lattice calculation for $T=140-155~\gev$.}
\label{fig:SphaleronStuff}
\end{figure}

The functions $\eta (T)$ and $\gamma (T)$ are defined in terms of the temperature $T$, the
temperature-dependent Higgs field expectation value $v_{min}$, and the Chern-Simons diffusion
rate $\Gamma_{diff} (T)$, plotted in Figure~\ref{fig:SphaleronStuff}.

\section{Results}

We assume $m_H \leq 2 m_X$, so that $H$ can go on-shell. We also assume the DM is in equilibrium up until $x=1$, after which we numerically solve the coupled Boltzmann/sphaleron equations.
As we only consider cases with $\langle \sigma_A v \rangle \gg \langle \sigma_{\bar X X \rightarrow \bar H L} v \rangle$, the annihilation process $\bar X X \rightarrow \bar H L, \bar L H$ will not significantly affect the WIMP miracle. Thus, the relevant parameters of our model are $m_X$, $m_H / m_X$, $\Gamma_H / m_H$ and $Re (\lambda_1 \lambda_2^*)$.

\begin{figure}[hear]
\center
\includegraphics[width=\textwidth]{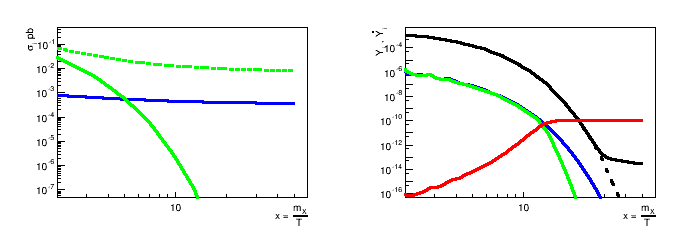}
\caption{The left panel shows the thermally-averaged cross sections $\langle \sigma_{XX}^{CP} v  \rangle$ (green dashed),
$\langle \sigma_{XX }^{CPV} v  \rangle$ (blue) and $\langle \sigma_{XL }^{CP} v  \rangle$ (green) (as defined
in eq.~\ref{eq:SimplifiedCrossSection}).
The right panel shows the corresponding contributions to the lepton source rate (sum of washout terms
in green and source term in blue) as well as $Y_B$ (red), $Y_X$ (black) and $Y_{X_{eq}}$ (black dashed).
We have chosen parameters
$ m_{X} = 5~\tev$, $ m_{H} = 7~\tev$,  $\lambda_1 = \lambda_2 = 0.5 $, $\langle \sigma_A v \rangle = 1~\pb$,
$\Gamma_H / m_H = 0.1 $ . }
\label{fig:injection}
\end{figure}

As scans over our parameter space are phenomenologically similar to~\cite{Bernal:2012}, we can distinguish our results with several benchmark points. For all benchmark models, the parameters are chosen to both generate the correct baryon asymmetry and the observed DM density.
In Figure~\ref{fig:injection}, we plot the thermally-averaged cross sections for the processes
$XX \rightarrow \phi^* \bar H' L$ (both $CP$-invariant and $CP$-violating terms) and
$XL \rightarrow \phi H' X$ (the $CP$-invariant term) as a function of $x = m_X / T$.
We also plot the contribution of these terms
to the lepton source injection rate, as well as $Y_B$, $Y_X$ and $Y_{X_{eq}}$.
We have chosen the ``high-mass" benchmark parameters
$m_{X} = 5~\tev$, $m_{H} = 7~\tev$, $\lambda_1 =  \lambda_2 = 0.5  $,
$\langle \sigma_A v \rangle = 1~\pb$, {\bf $\Gamma_H / m_H = 0.1$}.

Note that the $CP$-violating part of the $XX \rightarrow \phi^* \bar H' L$ process starts to drive the asymmetry as soon as the DM becomes non-relativistic, when $ x > 1 $. Although the departure from equilibrium is small at $ x \sim 1 $ relative to when DM freezes
out (typically $ x \sim 20 $), the DM density is much higher and even a small deviation from equilibrium will generate an asymmetry.
The asymmetry won't become large until washout processes freeze out, usually around $ x\sim 10 $ in WIMPy models~\cite{Bernal:2013bga}.

Other benchmarks include a narrower-width model with $m_{X} = 5~\tev$,
$m_{H} = 7~\tev$, $\lambda_1 =  \lambda_2 = 0.5 $, $\langle \sigma_A v \rangle = 1~\pb$, {\bf $\Gamma_H / m_H = 0.05$} 
and a low-mass model with $ m_{X} = 1.5~\tev$,  $m_{H} = 2.2~\tev$, $\lambda_1 =  \lambda_2 = 1$,
$\langle \sigma_A v \rangle = 1~\pb$, {\bf $\Gamma_H / m_H = 0.1$}.
These cases require stronger couplings in order to compensate for weaker asymmetry production.
The narrower-width benchmark is nearly identical to the high mass benchmark, only washout processes freeze out slightly later.
For the low-mass benchmark, sphalerons begin to decouple around when washout processes freeze out,
thus forcing a sharper freeze out of baryon number.
The parameters of these benchmark models are summarized in Table~\ref{tab:benchmarks}.

\begin{table}[hear]
\centering
\begin{tabular}{|c|c|c|c|c|c|c|}
\hline
benchmark & $m_X$ & $m_H$ &  $ \Gamma_H / m_H $  & $ \lambda_1 = \lambda_2 $ & $ \epsilon $ &
$ \langle \sigma_{XX \rightarrow \phi^* \bar H' L} v  \rangle  /  \langle \sigma_A v  \rangle $  \\
\hline
low-mass & $ 1.5~\tev $ & $ 2.2~\tev $ & $ 0.10 $ & $ 1.0 $ & $ 0.045 $  & $ 0.002 $ \\
\hline
high-mass & $ 5.0~\tev $ & $ 7.0~\tev $ & $ 0.10 $ & $ 0.5 $ & $ 0.045 $  & $ 0.008 $ \\
\hline
narrower-width & $ 5.0~\tev $ & $ 7.0~\tev $ & $ 0.05 $ & $ 1.0 $ & $ 0.022 $  & $ 0.033 $ \\
\hline
\end{tabular}
\caption{Benchmarks}
\label{tab:benchmarks}
\end{table}

The lack of ``pure" washout terms in our model allows us to generate the observed BAU with smaller $ \epsilon $ independently of
the mediator scale, $M_*$. We can then have $m_X = 1.5~\tev$ while still separating the UV physics from our new weak scale interactions. We also can have our $CP$-violating DM annihilation channel very small relative to any other channel one might consider in a more complete WIMP framework.

\section{Conclusion}

We have shown that the $CP$-violating process required by the Sakharov conditions for the dynamical generation of the BAU can arise
from tree-level diagrams and an absorptive final state interaction. This mechanism sequesters the one-loop suppression of the asymmetry and eliminates dangerous tree-level washout processes which are not Boltzmann suppressed. As a result, we can  effectively decouple our dark sector from the UV and address the DM relic density and BAU  more independently. We also
have carefully treated nonperturbative sphaleron effects through the electroweak crossover.

For simplicity, we have focused only on the approximation where the intermediate $H$ is a narrow resonance and can be produced on-shell.
Although one would expect an even more efficient asymmetry generation if we disregarded this assumption, additional washout processes incapable of producing a narrow resonance would have to be considered. Also, baryogenesis through DM annihilation directly into quarks would avoid the
sphaleron shutoff, but the mass of $H$ would be highly constrained by collider searches for DM production by quarks. In principal, the application of final state
absorptive interactions could also be used to eliminate dangerous washout processes from more traditional baryogenesis or leptogenesis scenarios.   

\Acknowledgements
J.~Kumar is coauthor of an article submitted for publication regarding this work, arXiv:1309.1145 [hep-ph].
We are grateful to S.~Pakvasa, X.~Tata, B.~Thomas, L.~Ubaldi, D.~Marfatia and B.~Garbrecht for useful discussions.
We thank the organizers of CosPA 2013 for their support and hospitality while this work was being completed.


\begin{thebibliography}{99}

\bibitem{WMAP:2011}
  {\bf WMAP} Collaboration, E.~Komatsu et al.,
  Astrophys J. Suppl. {\bf 192}, 18 (2011)
  [arXiv:1001.4538 [astro-ph.CO]].

\bibitem{Zurek:2013}
K.~M.~Zurek
  [arXiv:1308.0338 [hep-ph]].

\bibitem{Kohri:2009}
  K.~Kohri, A.~Mazumdar, N.~ Sahu and P.~Stephens,
  Phys.\ Rev.\ D {\bf 80}, 061302 (2009)
  [arXiv:0907.0622  [hep-ph]].

\bibitem{Cui:2012}
  Y.~Cui, L.~Randall and B.~Shuve,
  JHEP {\bf 1204}, 075 (2008)
  [arXiv:1112.2704 [hep-ph]].

\bibitem{Sakharov:1967dj}
  A.~D.~Sakharov,
  Pisma Zh.\ Eksp.\ Teor.\ Fiz.\  {\bf 5}, 32 (1967)
  [JETP Lett.\  {\bf 5}, 24 (1967\ SOPUA,34,392-393.1991\ UFNAA,161,61-64.1991)].

\bibitem{Bernal:2012}
  N.~Bernal, F.~Josse-Michaux and L.~Ubaldi,
  [arXiv:1210.0094 [hep-ph]].

\bibitem{Kumar:2013iva}
  J.~Kumar and D.~Marfatia,
  Phys.\  Rev.\ D {\bf 88}, 014035 (2013)
  [arXiv:1305.1611 [hep-ph]].

\bibitem{Kniehl:2008cj}
  B.~A.~Kniehl and A.~Sirlin,
  Phys.\ Rev.\ D {\bf 77}, 116012 (2008)
  [arXiv:0801.0669 [hep-th]].

\bibitem{Kolb:1990}
 E.~Kolb and M.~Turner,
   Front. Phys. {\bf 69}, 1-547 (1990).

\bibitem{D'Onofrio:2012}
  M.~D'Onofrio, K.~Rummukainen and A.~Tranberg,
  PoS(Lattice 2012) {\bf 055} (2012)
  [arXiv:1212.3206 [hep-ph]].

\bibitem{Burnier:2006}
  Y.~Burnier, M.~Laine and M.~Shaposhnikov,
  JCAP {\bf 0602}, 007 (2006)
  [arXiv:hep-ph/0511246].

\bibitem{Bernal:2013bga}
  N.~Bernal, S.~Colucci, F.~-X.~Josse-Michaux, J.~Racker and L.~Ubaldi,
  arXiv:1307.6878 [hep-ph].

\end{thebibliography}
\end{document}